\documentclass[aps,preprint,pra,showpacs]{revtex4}
\usepackage{graphicx}

\begin{document}

\title{A self-interfering clock as a ``which path" witness}

\author{Yair Margalit}	
	\affiliation{Department of Physics, Ben-Gurion University of the Negev, Beer-Sheva 84105, Israel}
\author{Zhifan Zhou}
	\affiliation{Department of Physics, Ben-Gurion University of the Negev, Beer-Sheva 84105, Israel}	
\author{Shimon Machluf}
\thanks{Present address: Van der Waals-Zeeman Institute, University of Amsterdam, Science Park 904, 1090 GL Amsterdam, The Netherlands}
	\affiliation{Department of Physics, Ben-Gurion University of the Negev, Beer-Sheva 84105, Israel}	
\author{Daniel Rohrlich}
	\affiliation{Department of Physics, Ben-Gurion University of the Negev, Beer-Sheva 84105, Israel}
\author{Yonathan Japha}
	\affiliation{Department of Physics, Ben-Gurion University of the Negev, Beer-Sheva 84105, Israel}	
\author{Ron Folman}
	\thanks{Corresponding author, email: folman@bgu.ac.il}
	\affiliation{Department of Physics, Ben-Gurion University of the Negev, Beer-Sheva 84105, Israel}

\date{\today}

\begin{abstract}
We experimentally demonstrate a new interferometry paradigm: a self-interfering clock. We split a clock into two spatially separated wave packets, and observe an interference pattern with a stable phase showing that the splitting was coherent, i.e., the clock was in two places simultaneously. We then make the clock wave packets ``tick" at different rates to simulate a proper time lag. The entanglement between the clock's time and its path yields ``which path" information, which affects the visibility of the clock's self-interference. By contrast, in standard interferometry, time cannot yield ``which path" information. As a clock we use an atom prepared in a superposition of two spin states. This first proof-of-principle experiment may have far-reaching implications for the study of time and general relativity and their impact on fundamental quantum effects such as decoherence and wave packet collapse.
\end{abstract}

\maketitle
Two-slit interferometry of quanta, such as photons and electrons, figured prominently in the Bohr-Einstein debates on the consistency of quantum theory \cite{BE,arr}.  A fundamental principle emerging from those debates---intimately related to the uncertainty principle---is that ``which path" information about the quanta passing through slits blocks their interference.  At the climax of the debates, Einstein claimed that a clock, emitting a photon at a precise time while being weighed on a spring scale to measure the change in its mass-energy, could evade the uncertainty principle.  Yet Bohr showed that the clock's gravitational redshift introduced enough uncertainty in the emission time to satisfy the uncertainty principle.  Inspired by the subtle role time may play in quantum mechanics, we have now sent a clock through a spatial interferometer.  The proof-of-principle experiment described below presents clock interferometry as a new tool for studying the interplay of general relativity \cite{GR} and quantum mechanics \cite{QM}.

Quantum mechanics cannot fully describe a self-interfering clock in a gravitational field.  If the paths of a clock through an interferometer have different heights, then general relativity predicts that the clock must ``tick" slower along the lower path. However, time in quantum mechanics is a global parameter, which cannot differ between paths. In standard interferometry (e.g. \cite{neutron}), a difference in height between two paths affects their relative phase and shifts their interference pattern; but in clock interferometry, a time differential between paths yields ``which path" information, degrading the visibility of the interference pattern \cite{Zych}. It follows that, while standard interferometry may probe general relativity \cite{IFM1,IFM2,IFM3}, clock interferometry probes the interplay of general relativity and quantum mechanics.  For example, loss of visibility due to a proper time lag would be evidence that gravitational effects contribute to decoherence and the emergence of a classical world---a world of events, such as measurement results---as predicted by R. Penrose \cite{penrose}, L. Diosi \cite{diosi} and others.

In our experiment, atomic clocks---atoms in superpositions of internal states---pass through an atomic matter-wave interferometer.
We demonstrate that the visibility of interference patterns produced by thousands of self-interfering clocks (atoms in a Bose-Einstein condensate) depends on the (simulated) proper time differential between the recombined wave packets of each clock. We simulate the time differential or lag by artificially making one clock wave packet ``tick" faster than the other. While our clock is not accurate enough to be sensitive to special- or general-relativistic effects, it is able to demonstrate that the proposal of Zych et al. \cite{Zych} is sound, namely, that a differential time reading affects the visibility of a clock self-interference pattern; specifically, the visibility equals the scalar product of the interfering clock states.

In principle, any system evolving with a well defined period can be a clock.  In our experiment, we utilize a quantum two-level system. Specifically, each clock is a $^{87}$Rb atom in a superposition of two Zeeman sublevels, the $m_F =1$ and $m_F=2$ sublevels of the $F=2$ hyperfine state.

\begin{figure}[t!]
      \centering
      \includegraphics[width=1\textwidth]{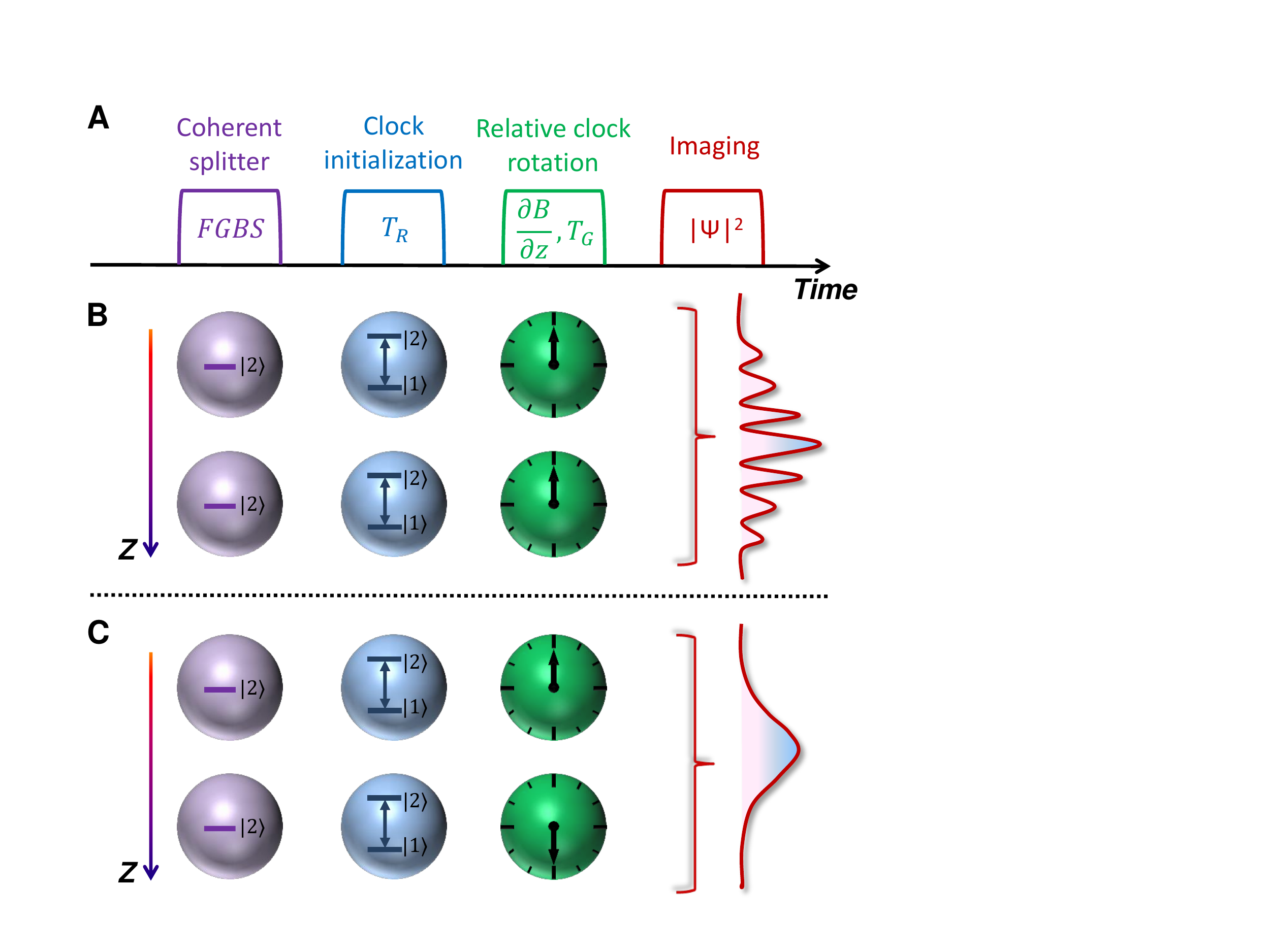}		
   \vskip-0.5\baselineskip
   \caption{(Color online)  Experimental sequence of the clock interferometer. (A) Detailed sequence (not to scale): Following a coherent spatial splitting by the FGBS and a stopping pulse, the system consists of two wave packets in the $|$2$\rangle$ state (separated in the direction of gravity, $z$) with zero relative velocity \cite{fgbs}. The clock is then initialized with an RF pulse of length $T_R$ after which the relative ``tick" rate of the two clock wave packets may be changed by applying a magnetic field gradient $\partial B / \partial z$ of duration $T_G$. Finally, before an image is taken (in the $xz$ plane), the wave packets are allowed to expand  and overlap for 8 ms. See \cite{SM} for a detailed description of the sequence.
   (B) Each clock wave packet shows as a one-handed clock, where the hand corresponds to a vector in the equatorial plane of the Bloch sphere. When the clock reading (i.e. the position of the clock hand) in the two clock wave packets is the same, fringe visibility is high. (C) When the clock reading is opposite (orthogonal), it becomes a ``which path" witness, and there is no interference pattern.  }
   \label{fig1}
\end{figure}

The general scheme of the clock interferometer is shown in Fig.\,1 (for additional information see \cite{SM}). To prepare the clock in a spatial superposition of two different locations, we make use of the previously demonstrated Stern-Gerlach type of matter-wave interferometer on an atom chip, creating a coherent spatial superposition of a $^{87}$Rb BEC \cite{fgbs} (about $10^4$ atoms $90~\mu$m below the chip surface). Initially, after the application of a field gradient beam splitter (FGBS) and a stopping pulse which zeroes the relative velocity of the two atomic wave packets, the wave packets are in the same internal atomic state ($|F,m_{F}$$\rangle$ = $|$2,2$\rangle$ $\equiv$ $|$2$\rangle$) as well as in the same external momentum state. The system's external wave function is thus $\psi(x-x_1)+\psi(x-x_2)$, where $x_i$ (i=1,2) are the mean values of the position of the two wave packets, which have the same center-of-mass momentum. A radio-frequency (RF) $\pi/2$ pulse (Rabi frequency $\Omega_R$ and duration $T_R$) tuned to the transition from $|$2$\rangle$ to $|$1$\rangle \equiv |$2,1$\rangle$ forms the clock by transferring the atoms from the $|$2$\rangle$ state to the internal superposition state ($|$1$\rangle$ + $|$2$\rangle$)/$\sqrt{2}$. The pulse is applied under a strong homogeneous magnetic field ($\bigtriangleup$E$_{12}$ $\approx h\times$25\,MHz) to push the transition to $|$2,0$\rangle$ out of resonance by $\sim$180\,kHz via the non-linear Zeeman effect, thus forming a pure two-level system for the $|$1$\rangle$ and $|$2$\rangle$ states.

\begin{figure}[t!]
      \centering
      \includegraphics[width=1.0\textwidth]{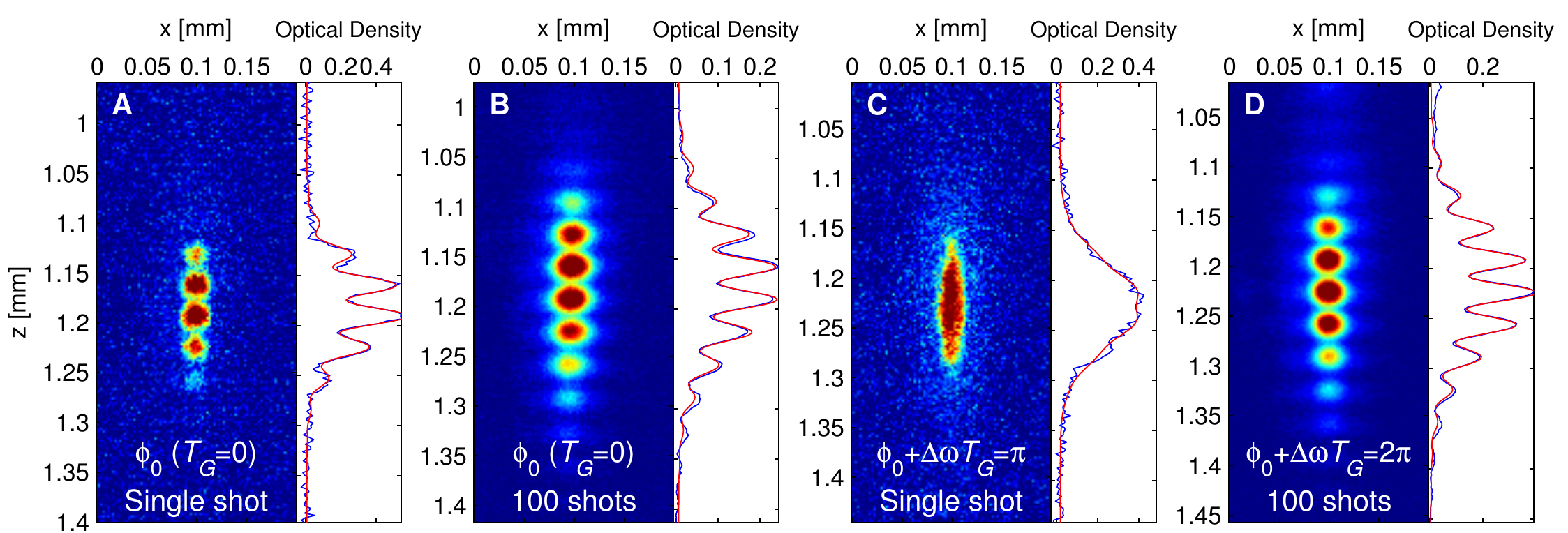}		
   \vskip-0.5\baselineskip
   \caption{(Color online) Clock interference: (A) A single experimental shot of a clock interfering with itself ($z$ axis values are relative to the chip surface). As $T_G=0$ the clock rate is approximately the same in the two wave packets and interference is visible. As can be seen from Fig. 3, a constant differential rotation of the clocks, $\phi_0$, exists even for $T_G=0$ (due to a residual magnetic gradient in our chamber \cite{SM}). This somewhat reduces the visibility. (B) To prove the coherence of the clock spatial splitting, an average of 100 consecutive shots such as that in (A) is presented, with only a minor change of visibility ($44\pm1\%$ compared to $46\pm4\%$ for the mean of the single-shot visibility \cite{SM}). (C) To prove that clock time acts as a ``which path" witness, we present a single shot in which the differential rotation angle $\phi_0+\Delta\omega T_G$ equals $\pi$. Unlike standard interferometers in which a phase difference does not suppress visibility, and contrary to standard split-BEC interference experiments in which a single shot always exhibits significant visibility, here the visibility is completely suppressed. (D) Similar to (B), but where one clock wave packet has been rotated by $2\pi$ relative to the other so that their readings are again indistinguishable (visibility is $47\pm1\%$, down from a single shot average of $51\pm2\%$). The fits are a simple combination of a sine with a Gaussian envelope \cite{fgbs}. Throughout this work, all data samples are from consecutive measurements without any post-selection or post-correction.}
   \label{fig2}
\end{figure}

In order to examine the coherence of the clock spatial superposition, we let the two clock wave packets freely expand and overlap to create spatial interference fringes, as shown in Fig.\,2(A). As two BEC wave packets are always expected to yield fringes when they overlap, many experimental cycles are required in order to prove phase stability or in other words coherent splitting of the clock. Fig.\,2(B) presents the averaged picture of $100$ single shots taken continuously over a period of about two hours. Relative to the mean of the single-shot visibility, the contrast falls by a mere 4$\%$, demonstrating a stable phase. The phase distribution in the data \cite{SM} reveals that the chance that the clock splitting is not coherent is negligible. We have thus proven with a high level of confidence that the clock has indeed been in two places at the same time.

We now show that clock time is indeed a ``which path" witness. For a single-internal-state interferometer, a phase difference will not change the visibility of the fringes. By contrast, the relative rotation between the two clock wave packets is expected to influence the interferometric visibility. In the extreme case, when the two clock states are orthogonal, e.g. one in the state of ($|1\rangle + |2\rangle$)/$\sqrt{2}$ and the other in the state of ($|1\rangle - |2\rangle$)/$\sqrt{2}$, the visibility of the clock self-interference should drop to zero. We therefore apply a magnetic gradient pulse (inducing a ``tick" rate difference $\Delta\omega$ \cite{SM}) of duration $T_G$ to induce a relative angle of rotation between the two clock wave packets (Fig.\,1). When the relative rotation angle is $\pi$, we observe in Fig. 2(C), in a single shot, that the visibility of the interference pattern drops to zero. Fig.\,2(D) exhibits a revival of the single-shot visibility when the differential rotation angle is taken to be $2\pi$ (where we again present an average of 100 shots to confirm coherence). It should be noted that the differential forces induced by the gradient are not strong enough to break the clock apart \cite{SM}.

To obtain a more general view of the effect, we present in Fig. 3 the dependence of the interferometer visibility on the differential rotation angle between the two clock wave packets over the range 0 to $4\pi$, by varying $T_G$ to alternate between clock indistinguishability and orthogonality---providing ``which path" information. The blue data present the clock interference pattern visibility, clearly showing oscillations (consistent with the expected period \cite{SM}). Comparing the latter oscillations to the visibility of a single-internal-state ``no clock" interference ($\Omega_R T_R=0$; red data) confirms that the oscillations are due to the existence of a clock.  The single-internal-state interference data also confirm that the overall drop in visibility is not due to the formation of the clock. This upper bound is due to the magnetic gradient pulse causing imperfect overlap between the two wave packets \cite{SM}. A lower bound on the visibility is due to the spatial separation of the $|1\rangle$ and $|2\rangle$ wave packets (i.e. gradual breakup of the clock), again due to the magnetic gradient \cite{SM}, which results in an increase of the visibility as expected from two independent single-state interferometers \cite{bounds}.

\begin{figure}[t!]
      \centering
      \includegraphics[width=0.6\textwidth]{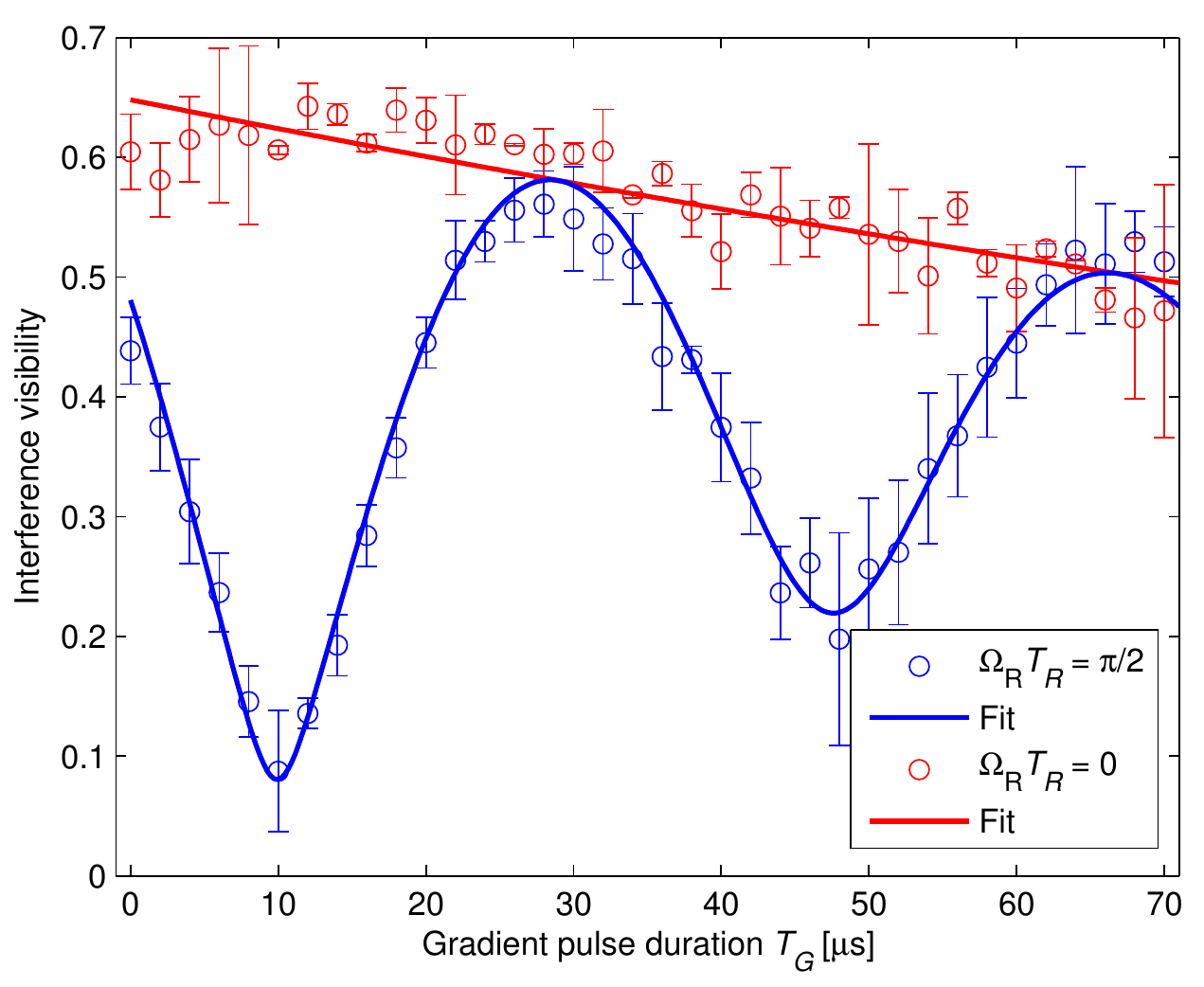}		
   \vskip-0.5\baselineskip
   \caption{(Color online) Varying the orthogonality of the two clock wave packets. To study the properties of clock time as a ``which path" witness, we measure visibility while continuously varying the relative rotation of the two clock wave packets (blue).
   Each data point is an average of the single-shot visibility obtained in several experimental cycles, and the error bars are the variance in this sub-sample. A fit returns an oscillation constant of $\Delta\omega = 0.166 \pm 0.003$ rad/$\mu$sec, consistent with an independent estimate (for further details see \cite{SM}). As inferred from the single-internal-state ``no clock" interferometer (red line) the oscillations are due to the existence of a clock. Regarding the upper and lower bounds, see text.  We note that the maximal visibility is slightly different from that of Fig. 2(D) as the data here are from a different run. }
   \label{fig3}
\end{figure}

The essence of the clock is that it consists of a superposition of two levels, i.e. $|$1$\rangle$ and $|$2$\rangle$. In Fig.\,3, we chose to work with an equal population of the $|$1$\rangle$ and $|$2$\rangle$ states upon clock initialization to create a proper clock, thus maximizing the visibility's dependence on the differential rotation. To further prove that it is the clock reading that is responsible for the observed oscillations of visibility, in Fig.\,4 we modulate the very formation of the clock by varying the clock-initiating RF pulse ($T_R$), so that the system preparation alternates between a proper clock and no clock at all. Specifically, varying $T_R$ changes the population ratio of the two components of the clock. When the differential rotation of the two clock wave packets ($\Delta\omega T_G$) is set to $\pi$ (i.e. orthogonal clocks), as shown by the blue data, the interferometer visibility oscillates as a function of the ratio of the clock states' initial population. This is so because when $\Omega_R T_R$ equals multiples of $\pi$ only one of the clock states is populated and the system is actually not a clock. In this case we have a standard interferometer; ``clock orthogonality" and clock time as a ``which path" witness do not exist irrespective of the fact that $\Delta\omega T_G=\pi$, and consequently full visibility is obtained. When a clock is formed (i.e. when the initial populations are similar), clock time is an effective witness, and the visibility drops. By contrast, when $\Delta\omega T_G=2\pi$ (red data), the interferometer visibility is always high because the two wave packets are not orthogonal whether they are clocks or states with a definite $m_F$.

\begin{figure}[t!]
      \centering
      \includegraphics[width=0.7\textwidth]{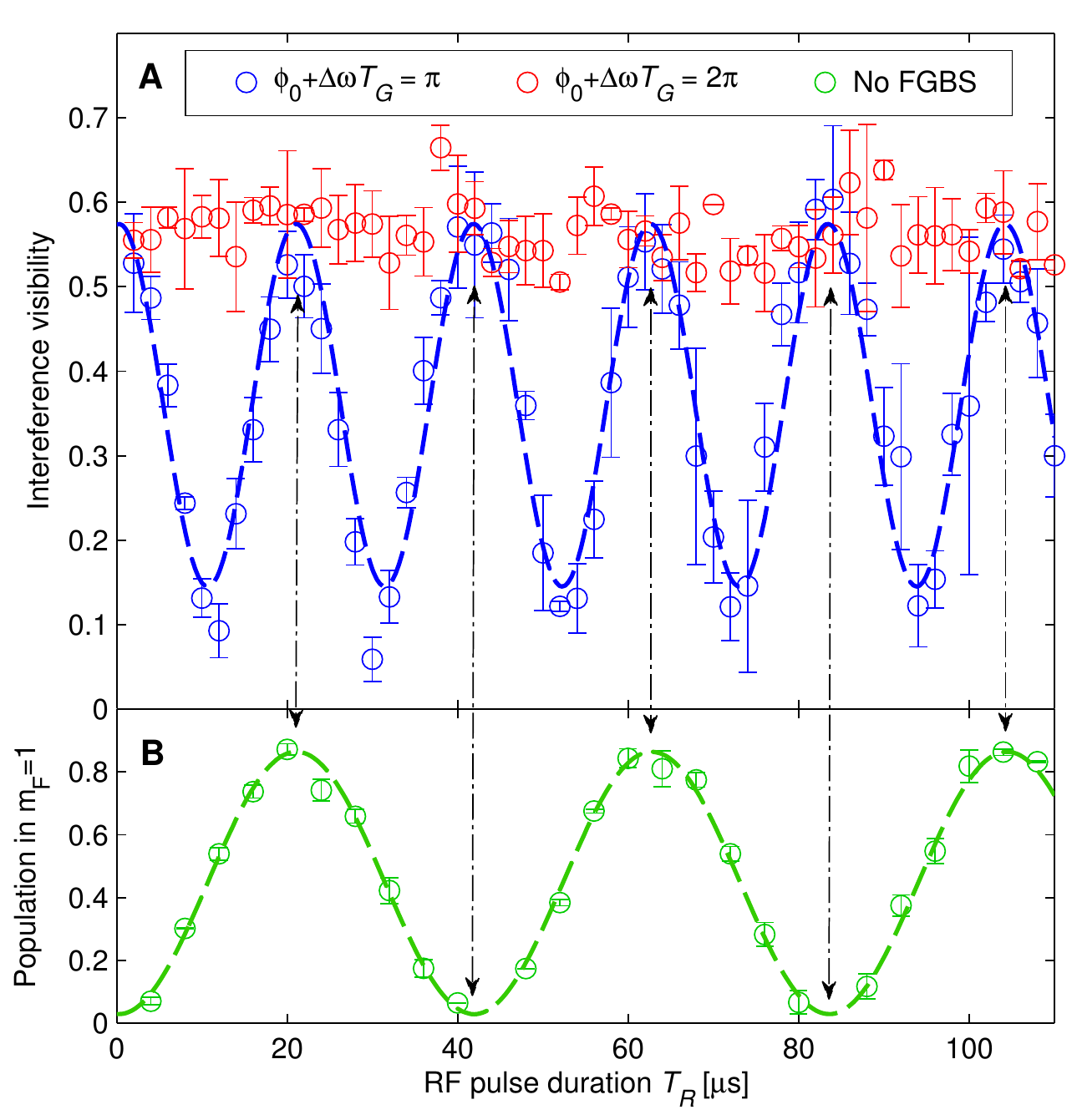}		
   \vskip-0.5\baselineskip
   \caption{(Color online) Varying the preparation of the clock. To further prove that it is the clock reading that is responsible for the observed oscillations in visibility, here we modulate the very formation of the clock by varying $T_R$, so that the system preparation alternates between a proper clock and no clock at all. (A) When the imprinted relative rotation between the two clock wave packets is $\pi$, whether a proper clock is formed or not has a dramatic effect (blue). By contrast, when the relative rotation is $2\pi$, whether a proper clock is formed or not has no effect (red). The error bars are the standard deviation of several data points. We note that in this measurement, there is no significant overall drop in visibility in the range of 5 oscillations as the RF pulse only changes the internal population. (B) The oscillation period appearing in (A) is as expected from an independent measurement of the Rabi oscillations induced by $T_R$ when the rest of the sequence has been eliminated \cite{SM}.}
   \label{fig4}
\end{figure}

Finally, we note that we consider recent works on the so-called Compton clock interferometer \cite{HM} and the debates that ensued (see \cite{ER,SP} and references therein) to be beyond the scope of this paper.

Future work will focus on clocks with increased accuracy and stability, first in the micro-wave and then in the optical regime, with the aim of reaching an accuracy allowing the experiment to detect relativistic effects. In addition, as time is considered by some a parameter which is still far from being fully understood \cite{TimeReBorn}, such an interferometer may shed new light on a variety of related fundamental questions.

\begin{acknowledgments}

We thank Zina Binstock for the electronics and the~BGU nano-fabrication facility for providing the high-quality chip. This work is funded in part by the Israeli Science Foundation, the~EC ``MatterWave'' consortium~(FP7-ICT-601180), and the German-Israeli~DIP project supported by the~DFG. We also acknowledge support from the~PBC program for outstanding postdoctoral researchers of the Israeli Council for Higher Education and from the Ministry of Immigrant Absorption (Israel).
D. R. thanks the John Templeton Foundation (Project ID 43297) and the Israel Science Foundation (grant no. 1190/13) for support. The opinions expressed in this publication do not necessarily reflect the views of the John Templeton Foundation.
\end{acknowledgments}

\newpage

\centerline{Supplementary material for}
\bigskip
\bigskip

\centerline{\bf A self-interfering clock as a ``which path" witness}

\bigskip

\centerline{Yair Margalit, Zhifan Zhou, Shimon Machluf, Daniel Rohrlich, Yonathan Japha, Ron Folman}

\centerline{\it Department of Physics, Ben-Gurion University of the Negev, Beer-Sheva 84105, Israel}

\bigskip
\bigskip

This file includes supplementary Sects. S1-S7, Figs. S1-S4, and Refs.
\cite{trad, bge, ar, extra1, extra2, extra3, extra4} (including Refs. \cite{extra1, extra2, extra3, extra4}, which propose cold-atom experiments on quantum gravity---or the effect of gravity---that are not directly related to our work, but included for completeness).


\bigskip

{\bf S1. Methods}

\newcommand{\1}{$|1\rangle$}
\newcommand{\2}{$|2\rangle$}
\newcommand{\bra}{\langle}
\newcommand{\ket}{\rangle}
\renewcommand{\vec}[1]{\boldsymbol{\mathbf{#1}}}
\newcommand{\phimag}{\phi_{\text{mag}}}
\newcommand{\mm}{\text{mm}}
\renewcommand{\theequation}{S\arabic{equation}}
\renewcommand{\thefigure}{S\arabic{figure}}
\setcounter{figure}{0}


Our experimental procedure is as follows. We begin by preparing a BEC of about $10^4$ $^{87}$Rb atoms in the state $\vert F, m_F\rangle =\vert 2,2\rangle$ in a magnetic trap located 90 ${\mu}$m below the chip surface. The trap is created by a copper structure located behind the chip. (See Fig. S1.) The BEC atoms are released from the trap, and 0.9 ms later a field gradient beam splitter (FGBS) \cite{fgbs} is applied to create a coherent spatial superposition of the BEC. The FGBS consists of one radio-frequency (RF) $\pi/2$ pulse (of 10 $\mu$s duration), a magnetic gradient pulse (4 $\mu$s), and another RF pulse (10 $\mu$s). These pulses create a superposition of $|2\rangle\equiv \vert 2,2\rangle$ wave packets having different momenta. After the FGBS, we apply a second magnetic gradient of 90 $\mu$s duration to zero the relative velocity of the wave packets. Clocks are initialized 1.5 ms after trap release by a third RF pulse of duration $T_R$, which creates a superposition of the $\vert 2\rangle$ and $\vert 1\rangle \equiv \vert 2,1\rangle$ states in each of the wave packets; then a phase-imprint magnetic gradient pulse of duration $T_G$ is applied in order to change the relative ``tick" rate of the two clock wave packets.

The entire sequence described above is done under a strong homogeneous magnetic bias field of 36.7 G in the $y$ direction, which creates an effective two-level system via the non-linear Zeeman effect. This field is adiabatically turned off 3.5 ms after the clock initialization (5 ms after the trap release), leaving earth's magnetic field to preserve the two-level system continuously. After an additional 3 ms time of flight (TOF)---thus 8 ms after the trap release---we image the atoms by absorption imaging and generate the pictures shown \hbox{in Fig. 2}.

\begin{figure}
\centerline{
\includegraphics*[width=\textwidth]{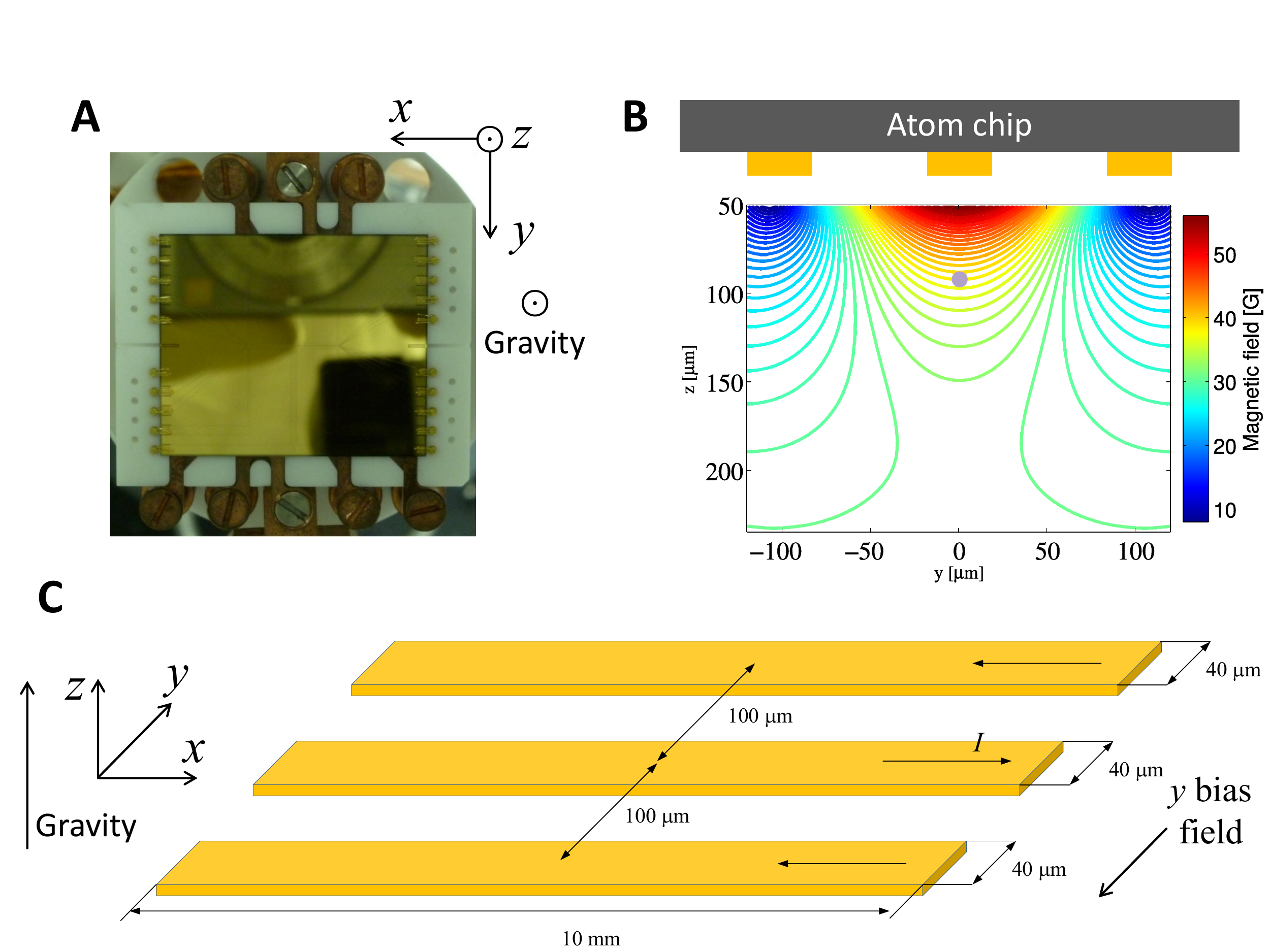}}
\caption{(A) A picture of the atom chip on its mount, with the copper structure visible behind it. Note that its orientation in the experimental setup is face down. (B) Magnetic field strength below the atom chip, generated by the quadrupole field via the chip wires and the bias field $B_y$ via external coils. The purple dot shows the location of the trapped BEC, which has in the $xz$ plane a half-width of about 9 $\mu$m. (C) Schematic diagram of the relevant chip wires. Wires are 10 mm long, 40 $\mu$m wide and 2 $\mu$m thick. The separation of the wires' centers is 100 $\mu$m, and the direction of the current $I$ alternates from one wire to the next.  The wires, being much smaller than the chip, are hardly visible in (A).}
\end{figure}

All three magnetic gradient pulses are generated by three parallel gold wires located on the chip surface (Fig. S1), which are 10 mm long, 40 $\mu$m wide and 2 $\mu$m thick.  The wires' centers are separated by 100 $\mu$m, and the same current runs through them in alternating directions, creating a 2D quadrupole field at $z$ = 100 $\mu$m below the atom chip. The FGBS phase noise is largely proportional to the magnitude of the magnetic field created during the gradient pulse \cite{fgbs}. As the main source of magnetic instability is in the gradient pulse originating from the chip, positioning the atoms near the middle (zero) of the quadrupole field created solely by the three chip wires 100 $\mu$m below the chip surface reduces the phase noise during the FGBS operation. The chip wire current was driven using a simple 12.5 V battery, and was modulated using a home-made current shutter, with ON/OFF times as short as 1 $\mu$s.  The total resistance of the three chip wires is 13.6 $\Omega$, yielding a current of 11.3/13.6 A $\approx 0.83$ A (a small voltage drop exists in the circuit itself).

The RF signal is generated by an Agilent 33250A waveform generator and subsequently amplified by a Minicircuit ZHL-3A amplifier.  We generate RF pulses using a Minicircuit ZYSWA-2-50DR RF switch.  RF radiation is transmitted through two of the copper wires located behind the chip (with their leads showing in Fig. S1).

\begin{figure}
\centerline{ \includegraphics[width=\textwidth]{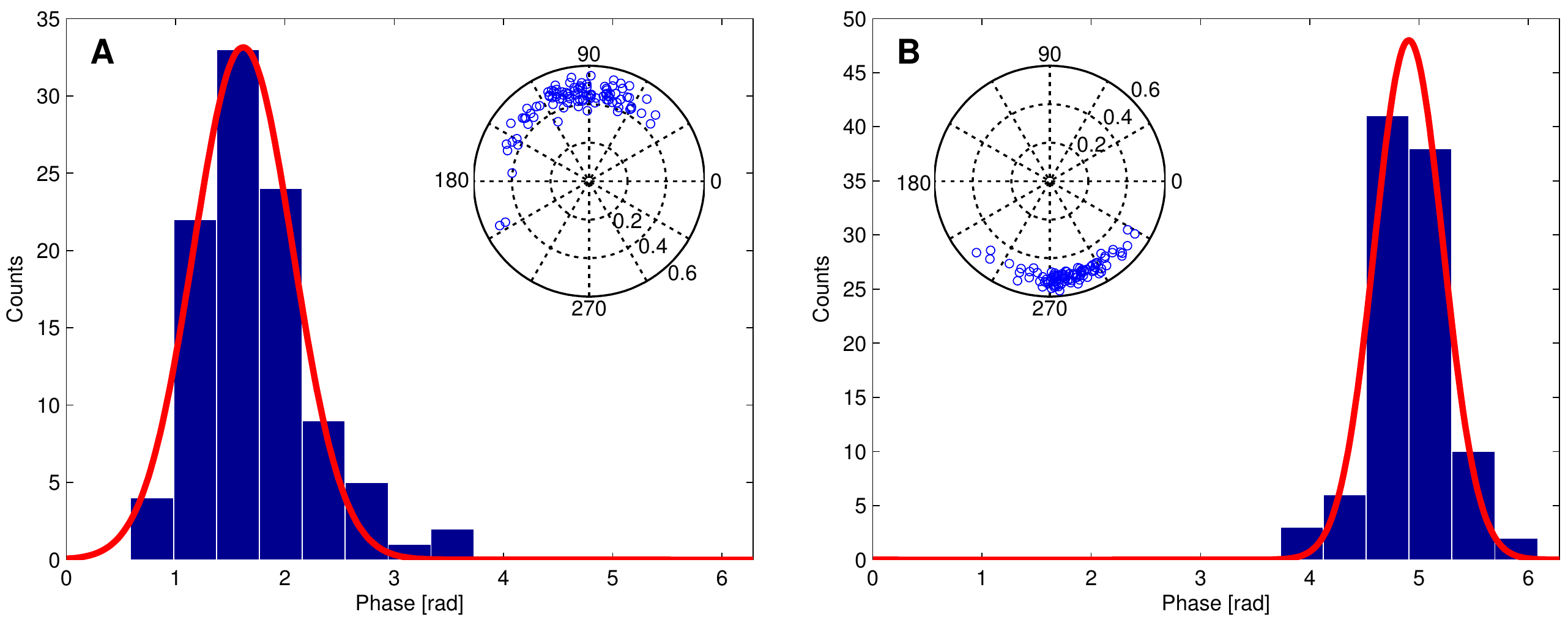}}\caption{Histograms of 100 shots used to generate the averaged pictures in Fig. 2. (A) Distribution of phases in Fig. 2(B), with a relative clock rotation of $\phi_0$ ($T_G=0$). (B) Distribution of phases in Fig. 2(D), with a relative clock rotation of $\phi_0+\Delta\omega T_G = 2\pi$. The standard deviations of the distributions are $\sigma = 0.454$ rad and $\sigma = 0.314$ rad for (A) and (B), respectively. Both distributions are significantly smaller than a random distribution of phases. Also shown are respective polar plots of phase vs. visibility (shown as angle vs. radius).}
\end{figure}

\begin{figure}
\centering{\includegraphics[width=0.5\textwidth]{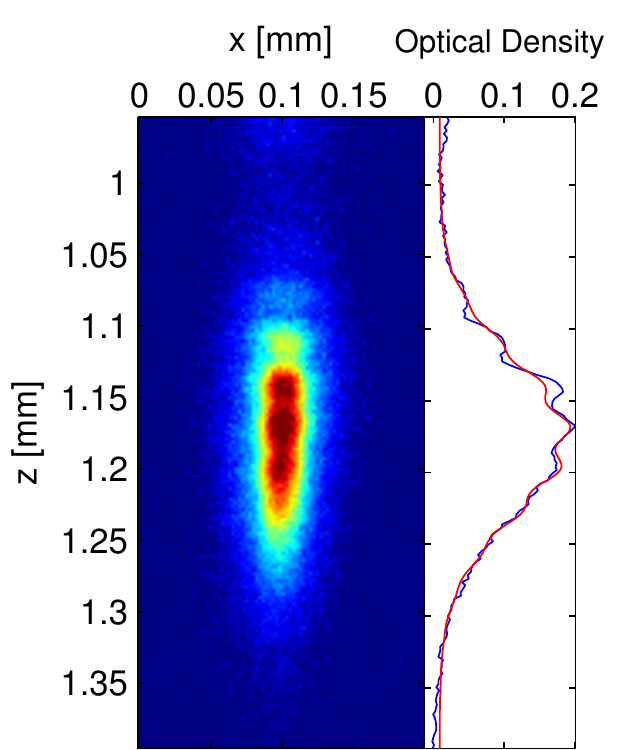}}\caption{An average of 65 consecutive single shots, in which the phase difference between the clocks is $\phi_0+\Delta\omega T_G = \pi$ as in Fig. 2C, yielding almost zero visibility. The minimal visibility ($\approx0.06$) that persists is consistent with the result of Fig. 3, in which a $\pi$ relative rotation yields a visibility of 0.09. Possible causes are the limited resolution (2 $\mu$s) in setting the time $T_G$, temporal jitter in the electronics that produces $T_G$, instabilities in the chip current, and a slight clock breakup due to the magnetic gradient.}
\end{figure}

\bigskip
{\bf S2. Coherence and visibility:  further analysis}

Fig. 2(B) shows a visibility of [44 $\pm$ 1]\% (error from fit). The mean visibility of 100 single shots is [46 $\pm$ 4] \% (error from standard deviation). The reduction of a mere 4\% [=(46-44)/46] from Fig. 2(A) to Fig. 2(B)  demonstrates a notable phase stability of the spatial interferometer.

Figs. S2(A-B) present the phase distributions of Figs. 2(B) and 2(D), each of 100 events. The probability that these phase distributions arise from a random phase process is negligible, which confirms the coherence of the clock interferometry.

Fig. S3 shows an average of 65 consecutive single shots of clock interference when the $\pi$ relative rotation is applied, yielding very low visibility ($\approx$ 0.06). This minimal but non-zero visibility is due partly to our limited time resolution in setting $T_G$, timing jitter of the electronics, and instabilities in the chip current.  See also the discussion of clock breakup in Sect. S6.

\bigskip
\goodbreak

{\bf S3. Fit and interpretation of Fig. 3}
\nobreak

The red data in Fig. 3 show a decay of the single-internal-state interference visibility as a function of $T_G$. We interpret this decay as the signature of non-perfect overlap of the two wave packets in space. Non-perfect wave-packet overlap can occur if the atoms are not located directly below the center of the three-wire structure, because of initial fluctuations in trapping position. When the magnetic gradients are applied, they cause the atoms to acquire some differential momentum along other axes in space beside the $z$ (gravity) direction, inducing some mismatch in the final $x$ and $y$ positions and thus reducing the interference visibility. Clearly this effect increases with $T_G$. In addition, the contrast of the clock visibility oscillations seen in Fig. 3 (blue data) is expected to decay due to the spatial separation of the $|2\rangle$ and $|1\rangle$ states within each of the two clock wave packets. This gradual breakup of the clock is due to the different force being applied during the magnetic gradient pulse on each of the states within the clock. It eventually results in an increase of the minimal visibility as orthogonality does not affect two non-overlapping single-state interferometers.

In order to understand the interference of the clock on the background of the overall decay of visibility with $T_G$, we first fit the single-internal-state interference data (red points in Fig. 3) to the expression $a\cdot e^{-T_G/\tau_1}$, where $a$ is the amplitude and $\tau_1$ is the corresponding decay rate. We then use these values of $a$ and $\tau_1$ in fitting the clock interference data (blue points in Fig. 3) to the expression $V(T_G)=ae^{-T_G/\tau_1}\sqrt{1-\sin^2[(\phi_0+\Delta\omega T_G)/2]/\cosh^2(\alpha_0+T_G/\tau_2)}$ (see Sect. S6), where $\alpha_0$, the initial decay of the contrast (of the clock's visibility oscillations), and $\tau_2$, the decay rate of the same contrast, are both related to the above-mentioned gradual breakup of the clock [the first related to a residual gradient in our system (see Fig. S4) and the second to the induced magnetic  gradient pulse, of duration $T_G$]; $\Delta \omega$ is the relative clock ``tick" rate; and $\phi_0$ is the relative rotation between the two clock wave packets, again caused by the residual magnetic gradient in our system.

An independent estimate of $\Delta\omega$ comes from calculating the expected magnetic field differences between the two wave packet positions during the magnetic gradient. From the Biot-Savart law, we obtain $B(z+\delta z)-B(z)=0.0387$ G. Here $\Delta \omega$ is
$0.17\pm 0.01 ~\rm{rad}/\mu \rm{s}$, where the uncertainty in $\Delta \omega$ is based on a position uncertainty of about 2 $\mu$m. This estimate is consistent with the fit result from Fig. 3 of $\Delta\omega =  0.166 \pm 0.003$ rad/$\mu$s.
\bigskip
\goodbreak

{\bf S4. Different clock preparations in Fig. 4}
\nobreak

While in Fig. 3 the time interval between the clock-initiating RF pulse ($T_R$) and the magnetic gradient ($T_G$) is minimal (8 $\mu$s), in Fig. 4 the gradient was delayed by 100 $\mu$s in order to allow $T_R$ to reach 100 $\mu$s and enable us to investigate different clock preparations.  Longer times make the atoms fall further away from the chip and the effect of the magnetic gradient produced by the chip diminishes.  Consequently $T_G$ was increased to produce the required phase shifts.  There is no significant decay observed in Fig. 4(A) as the RF pulse only changes the population ratio between $\vert 2\rangle$ and $\vert 1\rangle$, with no detrimental effects as a function of $T_R$, as can be seen in Fig. 4(B).

In order to measure Fig. 4(B) in the same position of the atoms as in Fig. 4(A), we cancel the RF pulses which are part of the FGBS and set $T_G=0$, while maintaining the magnetic-gradient pulses (FGBS and stopping pulse).  We then apply the clock-initiating RF pulse of duration $T_R$, and perform a Stern-Gerlach measurement of the relative population between $\vert 2\rangle$ and $\vert 1\rangle$. The Stern-Gerlach effect is created using a current pulse on the copper wires located behind the chip (Fig. S1).

\bigskip

{\bf S5. Effect of bias field inhomogeneity on the visibility}

As noted, our bias field is in the $y$ direction. Inhomogeneities of this field in the $z$ direction, along which the  two clock wave packets are separated, are expected to influence the relative ``tick" rate as well as the clock breakup.  In order to calculate their effect on the visibility of clock self-interference, we varied the duration of the bias field during the clock's time of flight (TOF) and measured the resulting visibility.  We extended the TOF by 10 ms (total TOF = 18 ms) in order to achieve longer clock times in the bias field. Fig. S4 shows that the visibility of clock self-interference depends on how long the clock is in the bias field. The effect is analogous to the oscillations seen in Fig. 3 (as a function of $T_G$), but here $\Delta\omega$ is much smaller since the inhomogeneity of the bias field is small. Hence, the small inhomogeneities in the bias field may be parameterized by $\alpha_0$ (Sect. S3) as causing an initial decay of the contrast in the visibility oscillations of the clock's interference pattern (due to clock breakup). These inhomogeneities also give rise to $\phi_0$ as they shift the differential clock ``tick" rate between the two clock wave packets.

\begin{figure}
\centering {\includegraphics[width=0.8\textwidth]{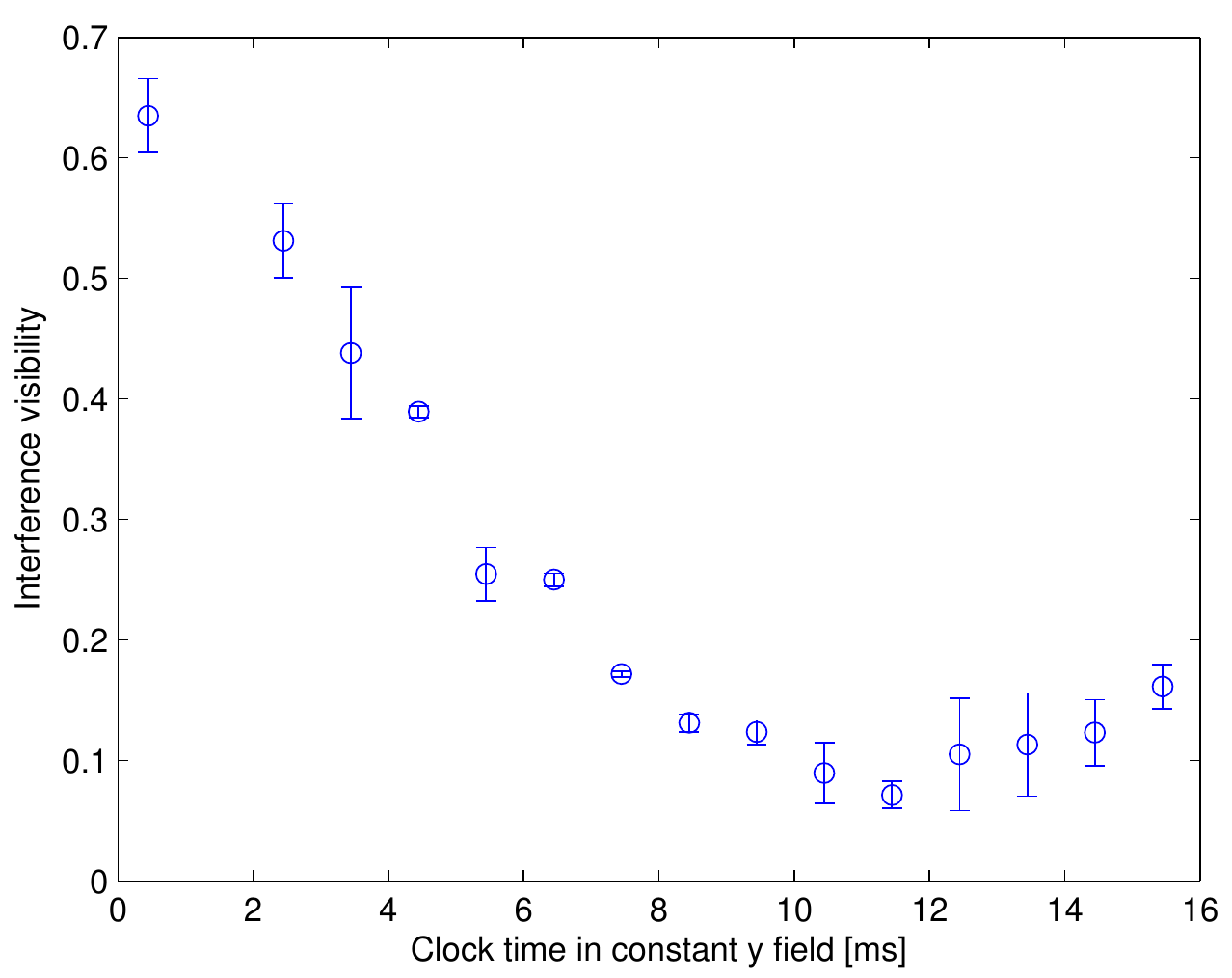}}\caption{Effect of inhomogeneities in the magnetic bias field on the visibility of clock self-interference. We vary the duration of the clock's stay in the bias field. The TOF is extended to 18 ms in order to achieve longer times in the bias field. The decrease and subsequent increase in visibility confirm that small inhomogeneities in the bias field induce the initial ($T_G=0$) reduction in visibility seen in Fig. 3, and are the cause of the initial relative phase $\phi_0$ of the clock wave packets. As the clock wave packets are separated vertically (in the $z$ direction), the effect is due to a residual gradient in the $z$ direction.}
\end{figure}

\bigskip
\goodbreak

{\bf S6. Model for the interference of the two clock wave packets}
\nobreak
\newcommand{\be}{\begin{equation}}
\newcommand{\ee}{\end{equation}}

To calculate the interference between two separated clock wave packets, we can model the wave packets
as Gaussian wave functions. We begin with
\be
e^{-\alpha(z-d/2)^2} \vert 2\rangle + e^{-\alpha(z+d/2)^2} \vert 2\rangle
\ee
as the state just after FGBS \cite{fgbs} and the stopping procedure; the state (corresponding to the
first stage in Fig. 1) is a superposition of two Gaussian wave packets with zero relative momentum separated by a distance $d$; here we neglect overall
normalization. Before the clock initialization, both wave packets are in the internal state $\vert 2\rangle$.
The parameter $\alpha$ is a complex number, and its real part ${\Re} [\alpha]= 1/2\sigma^2$ contains the width $\sigma$ of the Gaussian function. Its imaginary part is the coefficient of the quadratic phase due to the evolution of the wave packets during the preparation stage, especially the evolution due to harmonic components of the
magnetic potentials during the FGBS gradient pulse and the stopping pulse.
During free evolution, if $t_0>0$ this quadratic phase creates focusing and then divergence of the wave packet; or if $t_0 <0$, it creates divergence of the wave packet such that
its consequent evolution is similar to the evolution of a minimal wave packet from $t_0<0$. During free evolution of the wave packets, the parameter $\alpha$ evolves as
\be \alpha(t)=\frac{1}{2}[\sigma_0^2+i\hbar (t-t_0)/m]^{-1},
\ee
where $\sigma_0$ is the wave packet's minimal width at time $t=t_0$. The actual width $\sigma$ at time $t=0$
is related to these parameters by $\sigma=\sigma_0\sqrt{1+(\hbar t_0/m\sigma_0^2)^2}$.

Thus, at the end of the ``relative clock rotation" stage, which involves a magnetic field gradient $\partial B/\partial z$,
the state is
\be \left[e^{-\alpha(z-d/2)^2}+e^{-\alpha (z+d/2)^2}\right]\left(e^{ikz}\vert 1\rangle+
e^{2ikz}\vert 2\rangle\right),  \label{4terms} \ee
where $k=(\mu_B /2\hbar) \partial B /\partial z$, $\mu_B$ is the Bohr magneton, and $e^{ikz} \vert 1\rangle +e^{2ikz} \vert 2\rangle$ is the clock state. Note that the gradient pulse has two effects. First, it imprints different
phases $e^{ikd/2}$ and $e^{-ikd/2}$ on the two clocks centered at $z=d/2$ and $-d/2$, respectively. This relative phase is responsible for the ``which path" information that is carried by the clock. Second,
it applies a differential momentum kick $\hbar k$ and $2\hbar k$ to the two states $\vert 1\rangle$ and
$\vert 2\rangle$ that make up each clock state, leading to a partial breakup of the clock after propagation.
In what follows we examine quantitatively the effect of these two aspects of the gradient pulse on the
formation of the interference pattern.

Equation (\ref{4terms}) can be expressed as a sum of four terms that are equivalent up to exchanges $d \leftrightarrow -d$,
$\vert 1\rangle \leftrightarrow \vert 2\rangle$ and/or $k \leftrightarrow 2k$.  The time evolution
of each of the terms obeys the Schr\"odinger equation and the solution is well known:
\be\label{te}
e^{-\alpha(z\pm d/2)^2 +ikz}
\rightarrow e^{-i\hbar k^2 t/2m}
e^{-(z\pm d/2-\hbar kt/m)^2/[2(\sigma_0^2+i\hbar (t-t_0)/m)] +ik(z-\hbar k t/m)}
~~~~.
\ee
To exhibit the character of the interference pattern, it is sufficient that we consider the case
$\hbar (t-t_0)/m\gg \sigma_0^2$ (the far-field limit), in which the interference pattern is fully developed.
By detecting the atoms with an imaging light that does not distinguish between the two eigenstates
$\vert 1\rangle$ and $\vert 2\rangle$, one observes a sum of two fringe patterns with detection probability
$P(z)=|\langle 1|\psi\rangle|^2+|\langle 2|\psi\rangle|^2$, where $\vert \psi\rangle$ denotes the overall state. In a frame of reference moving with the center-of-mass momentum $3\hbar k/2$, it has the form
\begin{eqnarray}
P(z) &\propto &  \exp\left(-\frac{(z+\Delta z_t/2)^2}{\sigma_t^2}\right)
\cos^2\left[\frac{1}{2}\left(\frac{2\pi z}{\lambda_t}+\frac{\phi_t}{2}\right)\right] \cr
&&\cr
&& +\exp\left(-\frac{(z-\Delta z_t/2)^2}{\sigma_t^2}\right)
\cos^2\left[\frac{1}{2}\left(\frac{2\pi z}{\lambda_t}-\frac{\phi_t}{2}\right)\right]~~~,
\label{eq:twoexps}
\end{eqnarray}
where the two patterns have a periodicity $\lambda_t=2\pi\hbar(t-t_0)/md$ and an envelope width  $\sigma_t=\sigma_0\sqrt{1+(\hbar(t-t_0)/m\sigma_0^2)^2}$.
The two fringe patterns, one of state $\vert 1\rangle$ and the other of state $\vert 2\rangle$,  are shifted relative to one another by a phase $\phi_t=kdt/(t-t_0)$ and
their envelopes are shifted with respect to each other by $\Delta z_t=\hbar k t/m=\lambda_t\phi_t/2\pi$.
The shifts in phase and in space are one and the same and are both caused by the differential momentum kick applied at
the clock rotation stage. If the spatial shift $\Delta z_t$ is much smaller than the Gaussian width $\sigma_t$,
then the visibility of the joint fringe pattern depends on the
relative phase of the two fringe patterns in Eq.~(\ref{eq:twoexps}). If the relative phase $\phi_t$ is zero or
an even multiple of $\pi$, then the two interference patterns overlap and the visibility is 1. If $\phi_t$ is an odd
multiple of $\pi$ then the two fringe patterns are completely out of phase and the joint visibility drops to zero.
In the case where $\Delta z_t$ becomes considerable relative to the Gaussian width $\sigma_t$, each clock
wave packet is gradually broken into its constituents, eventually forming separate wave packets for the states
$|1\rangle$ and $|2\rangle$. In this case the cancelation of the two fringe patterns when $\phi_t=\pi$ is
not effective any more and the visibility does not drop to zero.

In order to examine the visibility more closely, we write Eq.~(\ref{eq:twoexps}) in a different form (omitting a constant pre-factor)
\begin{eqnarray} P(z) &\propto & e^{-z^2/\sigma_t^2}\left\{\cosh(z\Delta z_t/\sigma_t^2)
\left[1+\cos(\phi_t/2)\cos(2\pi z/\lambda_t)\right] \right. \nonumber \\
&&\left. +\sinh(z\Delta z_t/\sigma_t^2)\sin(\phi_t/2)\sin(2\pi z/\lambda_t)\right\}~~~~ .
\label{eq:singleexp}
\end{eqnarray}
This form can be further simplified to the form
\be P(z)\propto e^{-z^2/\sigma_t^2}\cosh(z\Delta z_t/\sigma_t^2)\left[1+{ V}(t,z)\cos[2\pi z/\lambda_t-
\varphi(t,z)\right], \ee
where
\be {V}(t,z)=\sqrt{1-\sin^2(\phi_t/2)/\cosh^2(z\Delta z_t/\sigma_t^2)} \ee
and
\be \varphi={\rm atan}[\tanh(z\Delta z_t/\sigma_t^2)\tan(\phi_t/2)]. \ee
If the breakup effect is small, namely $\Delta z_t\ll \sigma_t$, then
$\cosh(z\Delta z_t/\sigma_t^2)\sim 1$ and the visibility ${V}$ becomes  ${V}\approx |\cos(\phi_t/2)|$,
which is the scalar product of the two clock states $(|1\rangle+ e^{\pm i\phi_t/2}|2\rangle)/\sqrt{2}$.
In this case the phase $\varphi$ is also negligible. The clock breakup effect makes the visibility increase, such
that even when the clock states are completely orthogonal (i.e. distinguishable, when $\phi_t$ is an odd multiple of $\pi$) the visibility
drops to ${V}\approx \tanh(\Delta z_t/\sigma_t)$. In the intermediate range where the
clock breakup effect is significant but does not completely break the eigenstates apart, the phase $\varphi(t,z)$
causes  a $\phi_t$-dependent change of the periodicity of the fringes.

The effect of clock breakup follows from the fact that a constant gradient was applied in order to simulate the
time lag between the two clock wave packets. In this case a momentum kick to each wave packet is inevitable,
even if the two wave packets are well separated at the time when this clock rotation is applied.
In principle, the clock breakup effect could be avoided if the time lag could be formed by fields (magnetic or light) which have different
amplitudes at the position of the two clock wave packets but are constant across each of the wave packets.
\bigskip

{\bf S7. The clock interpretation}

\nobreak
We have interpreted our data in terms of a self-interfering clock.  Nevertheless it is instructive to discuss an alternate interpretation that does not refer to clocks at all. For the purpose of this discussion, we can simplify the interfering clock wave packets and write them as $\psi_+ (z)$ and $\psi_-(z)$, where
\begin{eqnarray}
\psi_+ (z) &=&
\psi (z) e^{i{\pi z/\lambda}}\left[ \vert 1\rangle+ e^{i\phi /2}~\vert 2\rangle \right]/2\cr
\psi_- (z) &=&
\psi (z) e^{-i{\pi z/\lambda}}\left[ \vert 1\rangle+ e^{-i\phi /2}~\vert 2\rangle \right]/2~~~,
\end{eqnarray}
and $\psi(z)$ is a (one-dimensional) localized, normalized wave function.  Then
$\psi_+ (z)$ and $\psi_-(z)$ represent counter-propagating and fully overlapping clock wave packets with a relative rotation angle $\phi$, and their sum corresponds to the interference of the two clock wave packets.  The visibility $V$ of the interference pattern depends on the relative clock angle $\phi$:
\begin{eqnarray}
\left\vert \psi_+ (z)+ \psi_- (z) \right\vert^2
&\equiv& \left[ \psi_+ (z)+ \psi_- (z) \right]^\dagger\left[ \psi_+ (z)+ \psi_- (z) \right]\cr
&&\cr
&=&{{\vert \psi (z)\vert^2}} \left[1+ \frac 1 4 e^{2\pi i z/\lambda} (1+e^{i\phi})
+\frac14 e^{-2\pi i z/\lambda} (1+e^{-i\phi})\right]~~~,
\label{vvp}
\end{eqnarray}
so $V=1$ for $\phi= 2n\pi$; for $\phi = (2n+1)\pi$ the clock states are orthogonal and the interference pattern is
\be
\left\vert \psi_+ (z)+ \psi_- (z) \right\vert^2=\vert \psi (z)\vert^2~~~,
\ee
with zero visibility.  In general, $V=\vert \cos (\phi /2)\vert$ and, as we have seen, $V=0$ only when the clock wave packets are orthogonal.  This calculation makes explicit the trade-off between the visibility of the interference and the ``which path" information arising from the clock \cite{KV1, KV2, trad, jea}.

However, we can rewrite $\left\vert \psi_+ (z)+ \psi_- (z)\right\vert^2$ by collecting terms in $\vert 1\rangle$ and $\vert 2 \rangle$.  We have
\be
\psi_+ (z)+ \psi_- (z) =
\psi(z)\left[ \vert 1 \rangle \cos ({\pi z/\lambda}) + \vert 2\rangle \cos ({\pi z/\lambda}+\phi /2)\right]~~~,
\ee
hence
\begin{equation}\label{phi}
\left\vert \psi_+ (z)+ \psi_- (z) \right\vert^2 = \vert \psi (z) \vert^2 \left[\cos^2 (\pi z/\lambda) + \cos^2 (\pi z/\lambda +\phi/2) \right]~~~,
\end{equation}
which represents {\it two} superimposed interference patterns, one from the $\vert 1 \rangle$ state and one from the $\vert 2\rangle$ state, with a relative phase that yields perfect visibility when $\phi =0$ (up to a multiple of $2\pi$) and zero visibility when $\phi = \pi$ (up to a multiple of $2\pi$), in agreement with Eq. \ref{vvp}, yet without any reference to a clock.

While this calculation is mathematically valid, it hides the physics of our experiment, in two ways.  First, nothing in this calculation refers to a clock.  There are innumerable physical realizations of a clock, and innumerable clock characteristics, such as accuracy and mass.  The clock in our experiment is the simplest possible---a superposition of two orthogonal states.  A superposition of $N > 2$ orthogonal states could be a much more accurate clock, and an external (rather than internal) variable could also serve to measure time \cite{ar}.  For any possible clock there would be a mathematical analysis analogous to the one above; for example, corresponding to a superposition of $N$ orthogonal states would be $N$ interference patterns that could add constructively to yield perfect visibility or destructively to produce a flat probability distribution.  What all these analyses would miss is the fact that the system analyzed is a clock (and therefore must measure proper time).

Second, this analysis fails to connect the loss of visibility to ``which path" information.  In Eq.\ (\ref{phi}) there is zero visibility when $\phi =\pi$, but it arises as the sum of two interference patterns, each with full visibility; since neither interference pattern is consistent with ``which path" information, how could we have guessed that their sum $is$ consistent with ``which path" information?

To conclude, among mathematical descriptions, the description via self-interfering clocks provides the best physical understanding of our experiment.


\begin{references}

\bibitem{BE} {N. Bohr, ``Discussions with Einstein on epistemological problems in atomic physics", in {\it Albert Einstein: Philosopher--Scientist}, ed. Paul A. Schilpp (New York:  Tudor Pub. Co.), 1951, pp. 201-41.

\bibitem{arr} Y. Aharonov and D. Rohrlich, {\it Quantum Paradoxes: Quantum Theory for the Perplexed} (Weinheim: Wiley-VCH), 2005, Sect. 2.4.}

\bibitem{GR} {``General relativity turns 100"}, Special issue, {\it Science} {\bf 347}, March 2015.

\bibitem{QM} {``Foundations of quantum mechanics"}, Special issue, {\it Nat. Phys.} {\bf 10}, April 2014.

\bibitem{neutron} R. Colella, A. W. Overhauser and S. A. Werner, ``Observation of gravitationally induced quantum interference", {\it Phys. Rev. Lett.} {\bf 34}, 1472 (1975).

\bibitem{Zych} {M. Zych, F. Costa, I. Pikovski and {\v{C}}. Brukner, ``Quantum interferometric visibility as a witness of general relativistic proper time", {\it Nat. Comm.} {\bf 2}, 505 (2011).}

\bibitem{IFM1} S. Dimopoulos, P. W. Graham, J. M. Hogan, and M. A. Kasevich, ``Testing general relativity with atom interferometry", {\it Phys. Rev. Lett.} {\bf 98}, 111102 (2007).

\bibitem{IFM2} H. M\"untinga et al., ``Interferometry with Bose-Einstein condensates in microgravity", {\it Phys. Rev. Lett.} {\bf 110}, 093602 (2013).

\bibitem{IFM3} C. C. N. Kuhn, et al., ``A Bose-condensed, simultaneous dual-species Mach-Zehnder atom interferometer", {\it New J. Phys.} {\bf 16}, 073035 (2014).

\bibitem{penrose}{R. Penrose, {\it The Emperor's New Mind:  Concerning Computers, Minds, and the Laws of Physics} (New York:  Oxford U. Press), 1989, Chap. 6.}

\bibitem{diosi}{L. Di\`osi, ``Gravity-related wave function collapse: mass density resolution", {\it J. Phys.: Conf. Ser.} {\bf 442}, 012001 (2013).}

\bibitem{SM}{Supplementary Material}

\bibitem{fgbs}{S. Machluf, Y. Japha and R. Folman, ``Coherent Stern-Gerlach momentum splitting on an atom chip", {\it Nat. Comm.} {\bf 4}, 2424 (2013).}

\bibitem{bounds} Between these two well understood bounds the ``which path" information \cite{KV1} or distinguishability \cite{KV2} (D), and visibility (V), should obey the relation $D^2+V^2\leq 1$.

\bibitem{KV1} P. D. D. Schwindt, P. G. Kwiat and B.-G. Englert, ``Quantitative wave-particle duality and nonerasing quantum erasure", {\it Phys. Rev.} {\bf A60}, 4285 (1999).

\bibitem{KV2} V. Jacques et al. ``Delayed-choice test of quantum complementarity with interfering single photons", {\it Phys. Rev. Lett.} {\bf 100}, 220402 (2008).

\bibitem{HM} S.-Y. Lan et al., ``A clock directly linking time to a particle's mass", {\it Science} {\bf 339}, 554 (2013).

\bibitem{ER} W. P. Schleich, D. N. Greenberger and E. M Rasel, ``Redshift controversy in atom interferometry: representation dependence of the origin of phase shift", {\it Phys. Rev. Lett.} {\bf 110}, 010401 (2013).

\bibitem{SP} S. Peil and C. R. Ekstrom, ``Analysis of atom-interferometer clocks", {\it Phys. Rev.} {\bf A89}, 014101 (2014).

\bibitem{TimeReBorn}{Lee Smolin, {\it Time Reborn} (Mariner Books), 2014.}

\bibitem{trad} {W. K. Wootters and W. H. Zurek, ``Complementarity in the double-slit experiment: quantum nonseparability and a quantitative statement of Bohr's principle", {\it Phys. Rev.} {\bf D19}, 473 (1979).

\bibitem{bge} B.-G. Englert, ``Fringe visibility and which-way information: an inequality", {\it Phys. Rev. Lett.} {\bf 77}, 2154 (1996).}

\bibitem{ar} {An example of an external variable as a measure of time appears in Y. Aharonov and D. Rohrlich, {\it op. cit.}, Sect. 8.5.}

\bibitem{extra1}  M. M. Dos Santos et al., ``Towards quantum gravity measurement by cold atoms", {\it J. Plasma Phys.} {\bf 79}, 437 (2013).

\bibitem{extra2} G. Amelino-Camelia et al., ``Constraining the energy-momentum dispersion relation with Planck-scale sensitivity using cold atoms", {\it Phys. Rev. Lett.} {\bf 103}, 171302 (2009).

\bibitem{extra3} D. E. Bruschi et al., ``Testing the effects of gravity and motion on quantum entanglement in space-based experiments", {\it New J. Phys.} {\bf 16}, 053041 (2014).

\bibitem{extra4} G. Rosi et al., ``Precision measurement of the Newtonian gravitational constant using cold atoms", {\it Nature} {\bf 510}, 518 (2014).


\end{references}
\end{document}